\documentclass[sigconf]{acmart}
\AtBeginDocument{%
  \providecommand\BibTeX{{%
    \normalfont B\kern-0.5em{\scshape i\kern-0.25em b}\kern-0.8em\TeX}}}

\setcopyright{acmcopyright}
\copyrightyear{2023}
\acmYear{2023}
\acmDOI{XXXXXXX.XXXXX}

\acmConference[Conference acronym 'XX]{Make sure to enter the correct
  conference title from your rights confirmation emai}{June 03--05,
  2023}{Woodstock, NY}
\acmISBN{978-1-4503-XXXX-X/18/06}

\usepackage{subfigure}
\usepackage{multirow}
\usepackage{bbding}
\usepackage{enumitem}
\usepackage{colortbl}
\usepackage{threeparttable}
\usepackage{paralist}

\begin{document}


\title{Rec4Ad: A Free Lunch to Mitigate Sample Selection Bias \\for Ads CTR Prediction in Taobao} 
\renewcommand{\shorttitle}{Rec4Ad: A Free Lunch to Mitigate Sample Selection Bias \\for Ads CTR Prediction in Taobao} 



\author{Jingyue Gao, Shuguang Han, Han Zhu, Siran Yang, Yuning Jiang, Jian Xu, Bo Zheng}
\email{{jingyue.gjy,shuguang.sh,zhuhan.zh,siran.ysr,mengzhu.jyn,xiyu.xj,bozheng}@alibaba-inc.com}
\affiliation{%
  \institution{Alibaba Group}
  \city{Beijing}
  \country{China}
}

\renewcommand{\authors}{Jingyue Gao, Shuguang Han, Han Zhu, Siran Yang, Yuning Jiang, Jian Xu, Bo Zheng}
\renewcommand{\shortauthors}{Gao, et al.}

\begin{abstract}
Click-Through Rate (CTR) prediction serves as a fundamental component in online advertising. A common practice is to train a CTR model on advertisement (ad) impressions with user feedback. Since ad impressions are purposely selected by the model itself, their distribution differs from the inference distribution and thus exhibits sample selection bias (SSB) that affects model performance. Existing studies on SSB mainly employ sample re-weighting techniques which suffer from high variance and poor model calibration. Another line of work relies on costly uniform data that is inadequate to train industrial models. Thus mitigating SSB in industrial models with a uniform-data-free framework is worth exploring. Fortunately, many platforms display mixed results of organic items (i.e., recommendations) and sponsored items (i.e., ads) to users, where impressions of ads and recommendations are selected by different systems but share the same user decision rationales. Based on the above characteristics, we propose to leverage recommendations samples as a free lunch to mitigate SSB for ads CTR model (Rec4Ad). After elaborating data augmentation, Rec4Ad learns disentangled representations with alignment and decorrelation modules for enhancement. When deployed in  Taobao display advertising system, Rec4Ad achieves substantial gains in key business metrics, with a lift of up to +6.6\% CTR and +2.9\% RPM.
\end{abstract} 



\maketitle

\section{Introduction}
For large-scale e-commerce platforms like Taobao, online advertising contributes a large portion of revenue. As advertisers typically pay for user clicks on advertisements (ads),  a common practice is to rank them based on expected Cost Per Mille (eCPM)~\cite{yuan2019improving}:
\begin{equation}
    eCPM = 1000 \times pctr \times bid,
\end{equation}\label{eq:ecpm}
where $pctr$ is the predicted Click-Through Rate (CTR), and $bid$ denotes the price for each click. Hence, CTR prediction serves as a fundamental component for online advertising systems. 

\begin{figure}[!htbp]
	\centering	
\includegraphics[width=0.65\columnwidth]{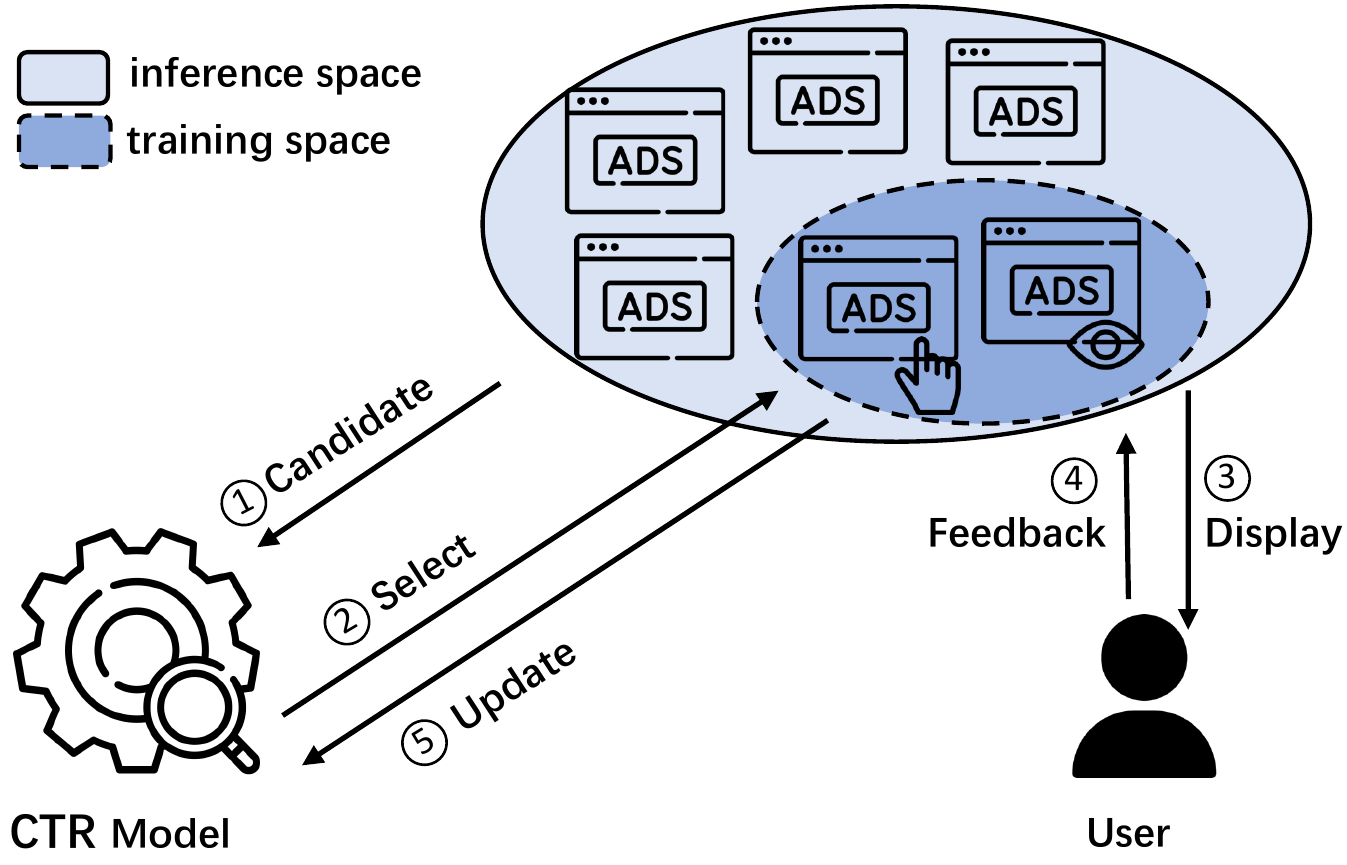}
    \caption{An illustration of the serving and updating process for CTR model in a typical online advertising system.}\label{fig:data-gene}
\end{figure}

As shown in Fig.~\ref{fig:data-gene}, a production CTR model scores all candidates and selects the top few based on Eq.~(\ref{eq:ecpm}) for display. The displayed ads as well as user feedback (i.e., click/non-click) are then recorded, with which we continuously train new models. Due to its simplicity and robustness, such a training paradigm is widely adopted by many industrial systems~\cite{anil2022factory,ma2022online,ling2017model}. However, since the displayed ads are not uniformly sampled from all candidates but purposely selected by the model itself, the training data distribution could be skewed from the inference distribution. This is widely known as the sample selection bias (SSB) problem ~\cite{zadrozny2004learning,heckman1979sample}. It violates the classical assumption of training-inference consistency and may potentially affect the model performance.

Recent efforts~\cite{wang2016learning,ovaisi2020correcting,pmlr-v48-schnabel16,uniform20,yuan2019improving} have been devoted to alleviating SSB in ranking systems. Methods based on Inverse Propensity Scoring~\cite{pmlr-v48-schnabel16,ovaisi2020correcting,wang2016learning} recover the underlying distribution by re-weighting the training samples. Despite theoretical soundness, they require a propensity model that accurately estimates sample occurrence probability, which is difficult to learn in dynamic and complicated environments. Moreover, sample re-weighting may yield un-calibrated predictions that are problematic for ads CTR models~\cite{yan2022scale}. Another line of work collects uniform data via random policy, which helps train an unbiased imputation model for non-displayed items~\cite{yuan2019improving} or guide the CTR model training via knowledge distillation~\cite{uniform20}. However, even small production traffic (e.g., 1\%) of the uniform policy will severely cause degraded user experience and revenue loss, and the obtained uniform data of this magnitude is insufficient for training industrial models with billions of parameters. With these issues, we investigate how to mitigate SSB for industrial CTR models under a \textbf{uniform-data-free} framework. 



\begin{figure}[!t]
	\centering	
	\includegraphics[width=0.85\columnwidth]{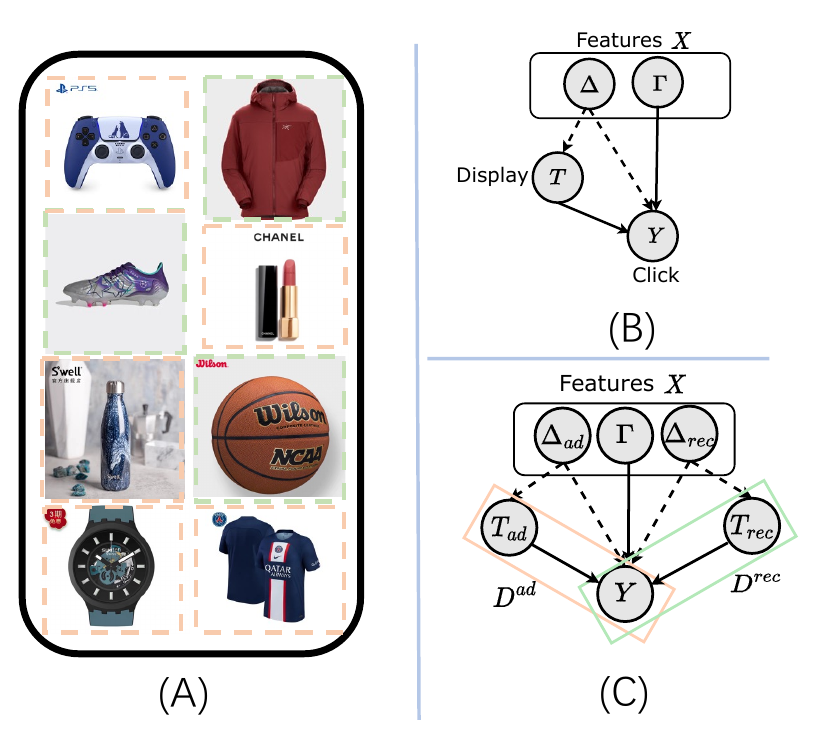}
    \caption{(A) Mixed results of ads (green dashed line) and recommendations (orange dashed line). (B,C) Causal Graph of CTR prediction for both ads and recommendations.}\label{fig:blended}
\end{figure}

Inspired by causal learning~\cite{wang2020causal,causE2018}, CTR prediction can be framed as the problem of treatment effect estimation. As in Fig.~\ref{fig:blended}(B), sample features compose \emph{unit} $\mathbf{X}$, whether to display it acts as a binary \emph{treatment} $\mathbf{T}$ and click is the \emph{outcome} $\mathbf{Y}$ to estimate. The root cause of SSB is attributed to existence of confounders $\mathbf{\Delta}$ (e.g., item popularity) in $\mathbf{X}$ that affect both $\mathbf{T}$  and $\mathbf{Y}$. Recent studies~\cite{hu2022improving,geirhos2018imagenettrained} show that confounders mislead models to capture spurious correlations between the unit features and the outcome, which are non-causal and hurt generalization over the inference distribution. Hence, it is promising to mitigate SSB by disentangling confounders $\mathbf{\Delta}$ from real user-item interest $\mathbf{\Gamma}$ in sample features, which is non-trivial in absence of randomized controlled trials (i.e., uniform data)~\cite{pearl2009causality}.


As shown in Fig.~\ref{fig:blended}, many platforms~\cite{chen2019learning,goldfarb2011online} display mixed results of sponsored items and organic items that are independently  selected by advertising and recommendation systems. For clarity, \textbf{we refer to sponsored items as ads}. \textbf{We refer to organic items as recommendations}. This scenario has two characteristics: 
\begin{compactitem}
    \item \textbf{Shared decision rationales.} With a unified interface design, users are unaware of whether items are sponsored or organic, making their click decision determined by real user-item interest $\mathbf{\Gamma}$ rather than sources of displayed items.
    \item \textbf{Different selection mechanisms.} Advertising and recommendation systems serve different business targets (e.g., revenue/clicks/dwell time)~\cite{goldfarb2011online} and have different selection mechanisms as verified in Sec.~\ref{sec:ana}. Thus their SSB-related confounders are rarely overlapped, making system-specific confounders $\mathbf{\Delta_{ad}}$/$\mathbf{\Delta_{rec}}$ capture a substantial portion of $\mathbf{\Delta}$.  
\end{compactitem}
 The above characteristics make it possible to disentangle $\mathbf{\Delta_{ad}}$/$\mathbf{\Delta_{rec}}$ and $\mathbf{\Gamma}$ by jointly considering samples from two sources. 
 Compared with the uniform data, recommendation samples are of a comparable or even larger magnitude than ads samples and persist without revenue loss, making it a free lunch worthy of exploitation. Though few if any confounders common in two systems could still remain with $\mathbf{\Gamma}$, disentangling system-specific $\mathbf{\Delta_{ad}}$/$\mathbf{\Delta_{rec}}$ from $\mathbf{\Gamma}$ is already a meaningful step towards mitigating SSB, especially when uniform data is unavailable in industrial advertising systems.

To this end, we propose to leverage \underline{\textbf{Rec}}ommendation samples to mitigate SSB \underline{\textbf{F}}or \underline{\textbf{Ad}}s CTR prediction ({\textbf{Rec4Ad}}). Under this framework, recommendation samples are retrieved and mixed with ad samples for training. With raw feature embeddings, we elaborately design the representation disentanglement mechanism to dissect system-specific confounders and system-invariant user-item interest across two systems. Specifically, this mechanism consists of an alignment module and a decorrelation module with various regularizations. Finally we make prediction with disentangled and enhanced representation. Rec4Ad has been deployed to serve the main traffic of Taobao display advertising system since July of 2022.

Our contributions are summarized as follows:
\begin{compactitem}
    \item We analyze the existence of SSB in CTR prediction and point out the potential to leverage recommendation samples to mitigate such bias in absence of uniform data.  
    
    \item We propose a novel framework named Rec4Ad, which jointly considers the recommendation and ads samples in learning disentangled representations that dissect system-specific confounders and system-invariant user-item interest. 
    
    
    \item We conduct offline and online experiments to validate the effectiveness of Rec4Ad that achieves substantial gains in business metrics (up to +6.6\% CTR and +2.9\% RPM). 
\end{compactitem}

\section{Preliminary}
\subsection{Problem Formulation}
\noindent\textbf{Input:} The input includes a user set $\mathbf{\mathcal{U}}$, an ad set $\mathbf{\mathcal{A}}$, an item set $\mathbf{\mathcal{I}}$, user-ad impressions $\mathbf{\mathcal{D}^{ad}}$, and user-item simpressions $\mathbf{\mathcal{D}^{rec}}$
\begin{compactitem}
    \item Each user $u \in \mathbf{\mathcal{U}}$ is represented by a set of features $\{u_1, ..., u_m\}$ including user profile features (e.g., age and gender) and historical behaviors (e.g., click and purchase).
    \item Each ad $a \in \mathbf{\mathcal{A}}$ is a promotion campaign for a sponsored item $i \in \mathbf{\mathcal{I}}$. Besides item-level features like category and brand, $a$ also has campaign-level features including ID and historical statistics, denoted by $\{a_1,...,a_n\}$. 
    \item Each impression in $D^{ad}$ is a tuple $(u,a,c,y)$ describing when the advertising system displayed $a$ to $u$ under context $c = \{c_1,..,c_k\}$ such as time and device, user clicked it ($y=1$) or not ($y=0$). As for $D^{rec}$, the tuple changes to $(u,i,c,y)$ logged by the recommendation engine.
\end{compactitem}

\noindent\textbf{Output:} We aim to learn a model $f$ that predicts the click probability $f(u,a,c)$ if displaying $a \in \mathbf{\mathcal{A}}$ to user $u \in \mathbf{\mathcal{U}}$ under context $c$.

\subsection{Analysis of Sample Selection Bias}\label{sec:ana}

\begin{figure*}[!htbp]
    \centering
    \includegraphics[width=0.88\textwidth]{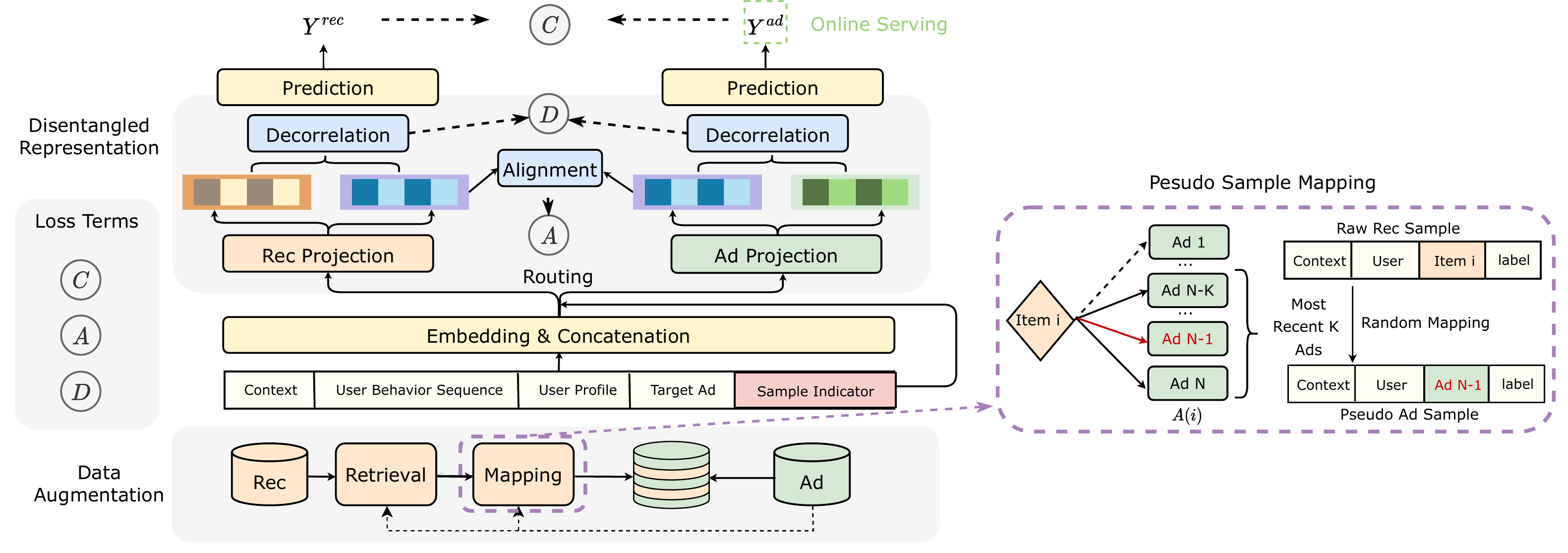}
    \caption{An overall framework of Rec4Ad (best viewed in color).}
    \label{fig:framework}
\end{figure*}

\begin{figure}[!htbp]
\centering	
\includegraphics[width=0.47\columnwidth]{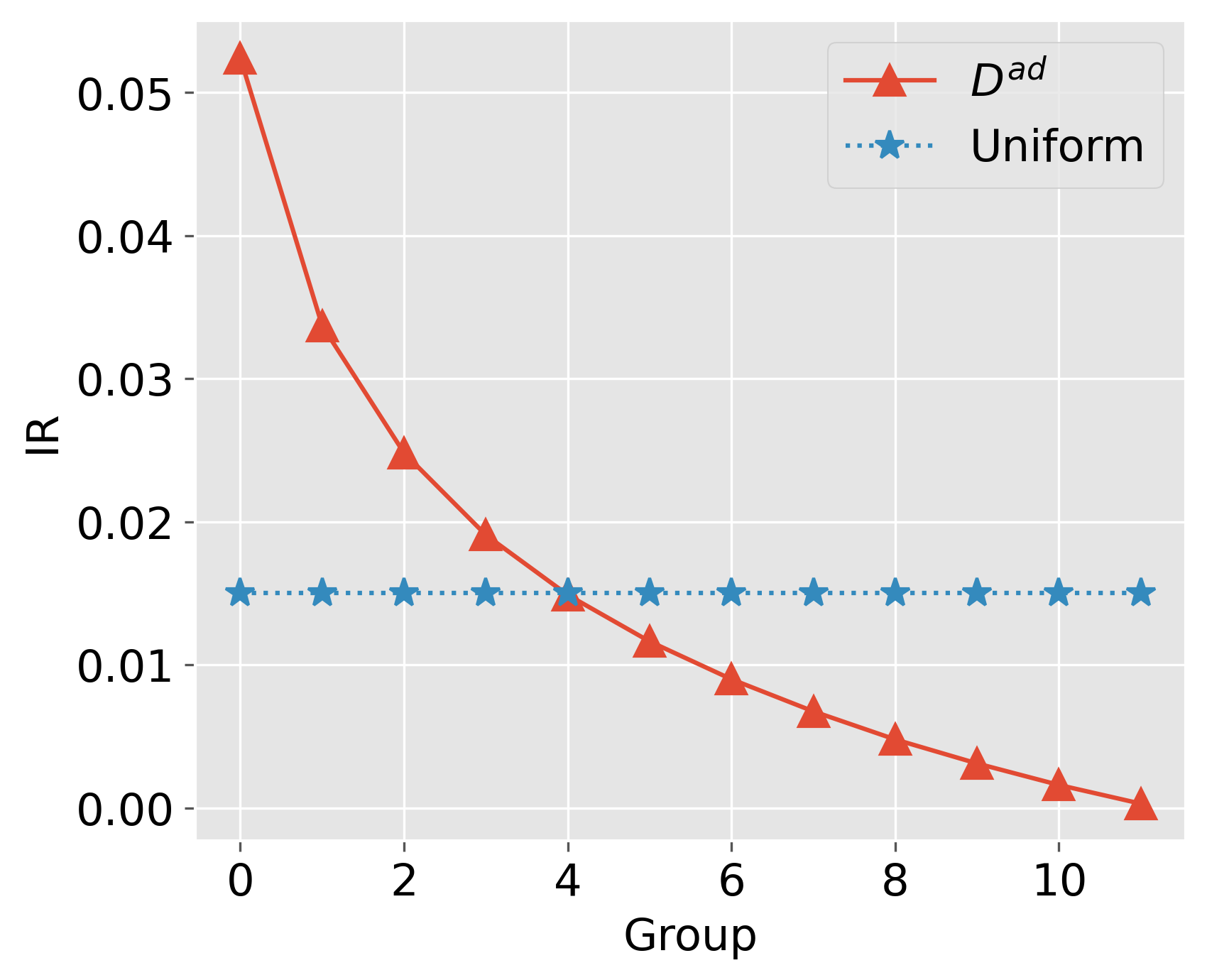}    
\includegraphics[width=0.52\columnwidth]{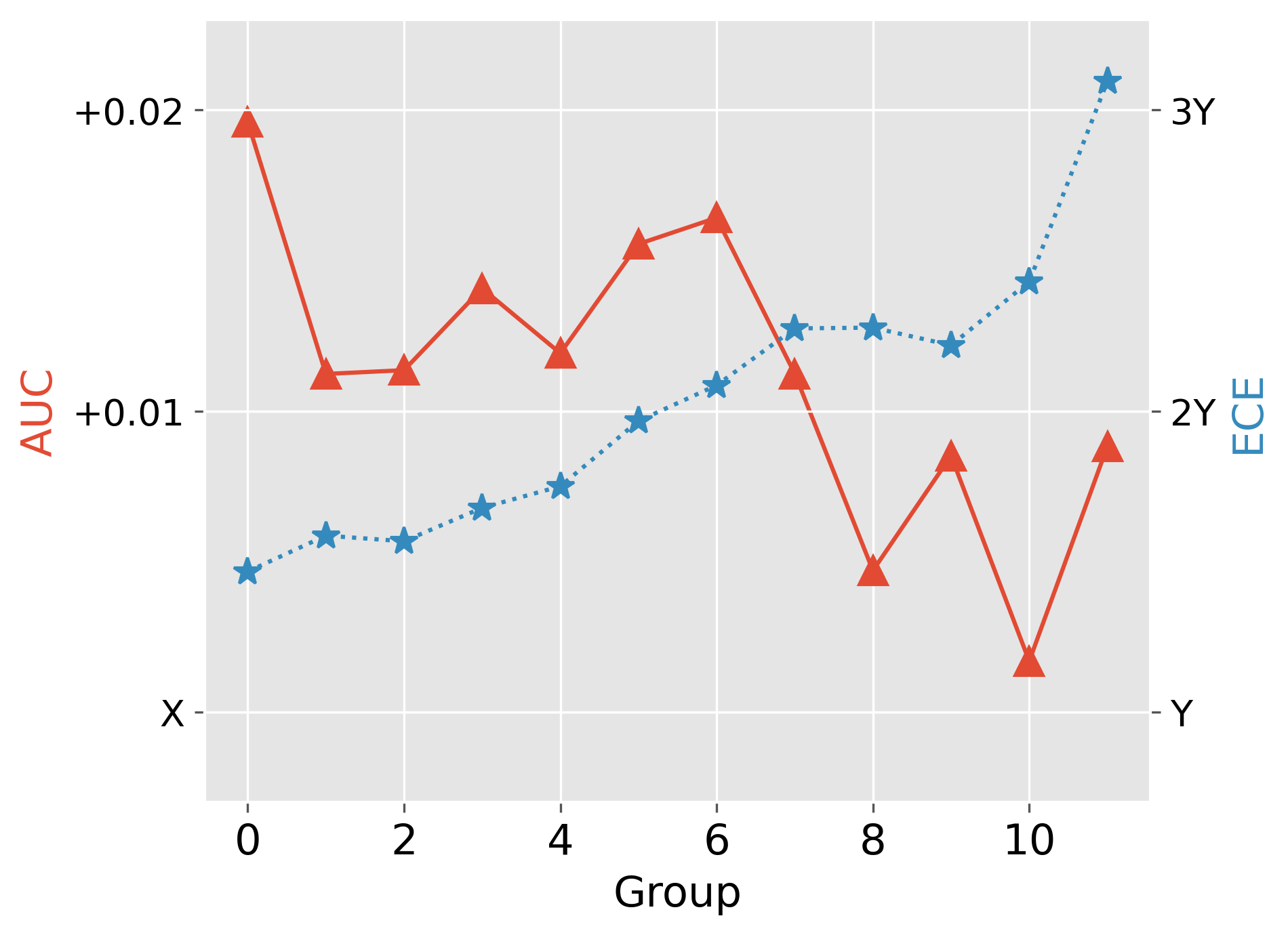}   
\caption{Left: Average IR for different groups of ads. Right: Model performance w.r.t different IR group of ads.}\label{fig:ir}
\end{figure}

\begin{figure}[!htbp]
	\centering	
	\includegraphics[width=0.48\columnwidth]{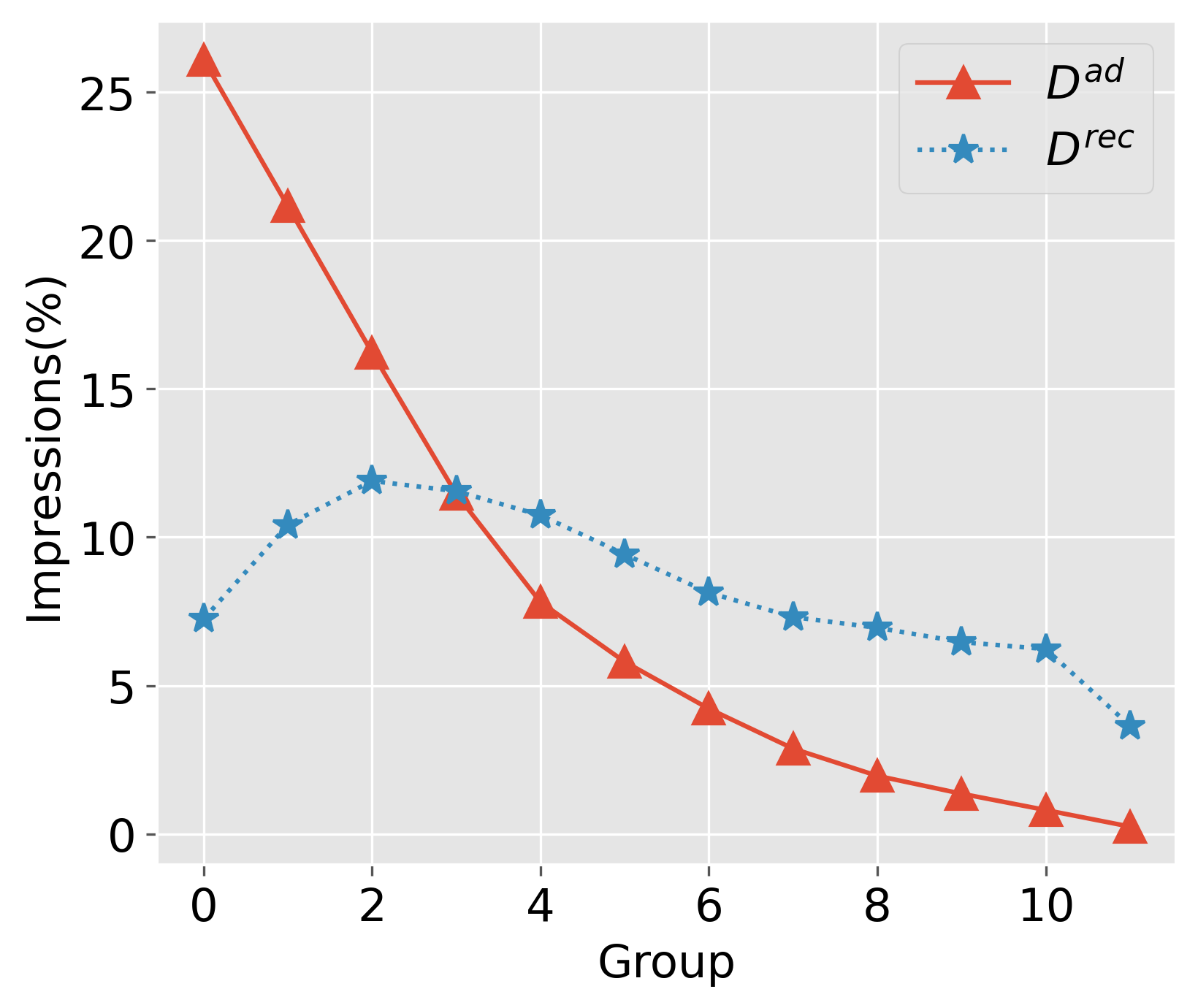}    
	\includegraphics[width=0.48\columnwidth]{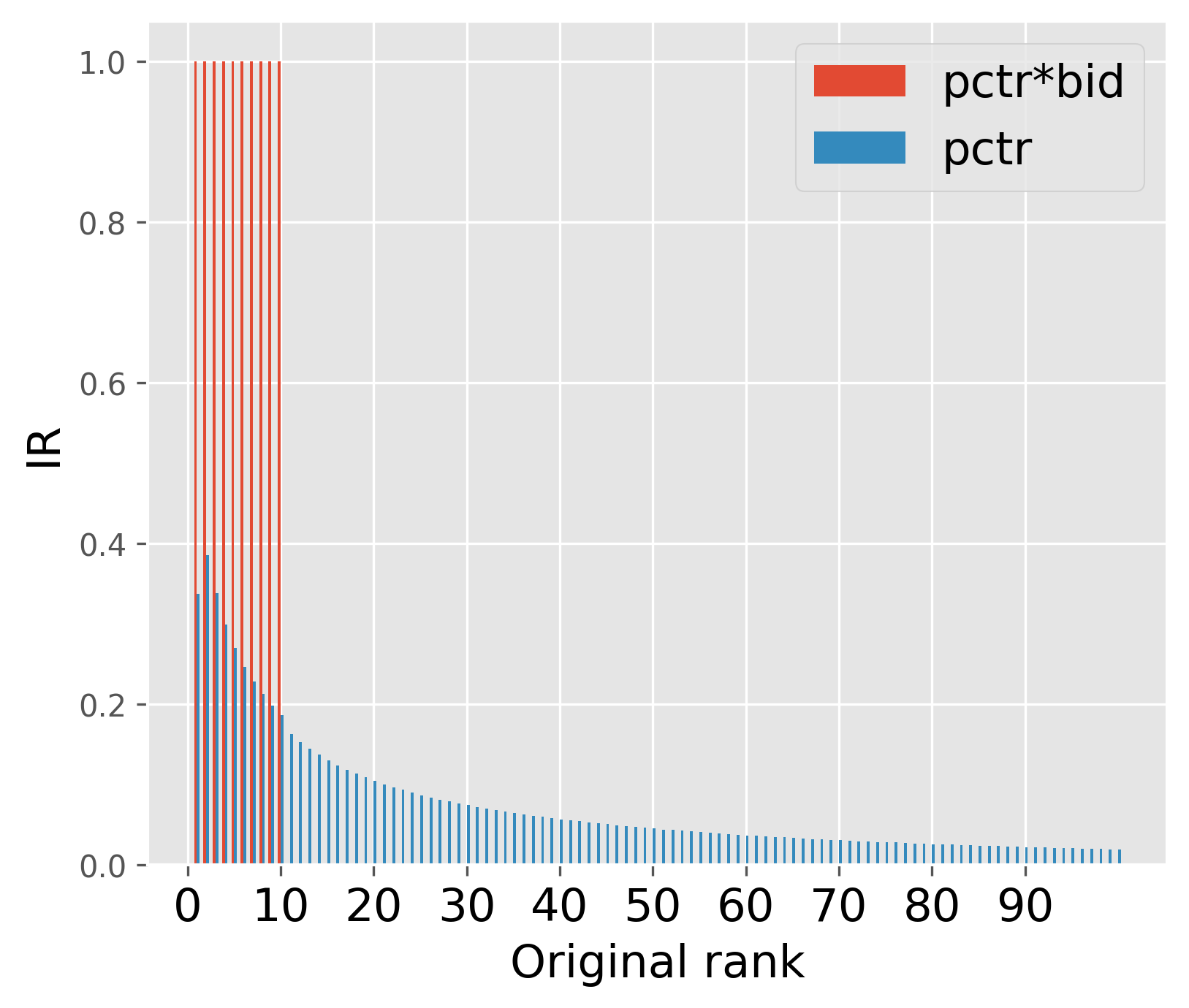}   
	\caption{Left: Impression Distribution in $D^{ad}$ and $D^{rec}$. Right: Simulated results under different ranking mechanisms.}\label{fig:rec_ad}
\end{figure}

\subsubsection{SSB in Ads Impressions}
SSB happens when each candidate does not share equal opportunities for impression. To examine its existence, we define the metric of \emph{impression ratio} (IR) to measure the opportunity of each ad in our system:
\begin{equation}\label{eq:ir}
    \textbf{IR}(a) = \frac{\#\text{sessions where }a\text{ was displayed}}{\#\text{sessions where }a\text{ was in the candidate set}},
\end{equation}
where a session refers to a user request. 
We first calculate IR for each ad in $D^{ad}$, sort them by IR in descending order, and then divide them equally into 12 groups. Fig.~\ref{fig:ir} (Left) shows the average IR for each group (the red line) compared with the ideal uniform data (the blue line). We find that \textbf{impressions on $D^{ad}$ are distributed among different ads in an extremely imbalanced way}, where the IR of the first group is nearly 200 times that of the last group. 


\subsubsection{Influence of SSB}\label{sec:influence}
Since base CTR model is trained on $D^{ad}$, we analyze its ranking and calibration performance under the imbalanced ad impressions. For ranking performance, we use the metric of AUC, and the calibration performance is measured by the Expected Calibration Error (ECE)~\cite{yan2022scale}. Details of two metrics are introduced in Sec.~\ref{sec:setup}. Fig.~\ref{fig:ir} (Right) shows online model performance on different groups of ads with descending IR. It is observed that \textbf{model tends to perform worse on ads with lower IR than on those with higher IR}. It is consistent with our assumption that model does not generalize well on ads with few impressions and validates the necessity to handle SSB for improvement.


\subsubsection{Mitigating SSB with Recommendation Samples.} As defined before, each ad in $\mathbf{\mathcal{A}}$ corresponds to an organic item in $\mathbf{\mathcal{I}}$. Thus we investigate how impressions in $D^{rec}$ are distributed among organic items with ad counterparts. From Fig.~\ref{fig:rec_ad} (Left), we find that \textbf{though some groups of ads have few impressions in $D^{ad}$, their corresponding items contribute an important portion of impressions in $D^{rec}$}. It is attributed to different selection mechanism behind $D^{ad}$ and $D^{rec}$. We conduct a simulated study on $D^{ad}$ for verification, which changes the ranking function from $pctr*bid$ to $pctr$ (commonly adopted in recommendation systems) and re-displays top-10 ads. Fig.~\ref{fig:rec_ad} (Right) illustrates IR of each original rank under two mechanisms. It is clear that \textbf{impression distribution is changed, where ads with low rank in original list have opportunities to be displayed under another mechanism}. Above empirical analysis show that it is promising to leverage recommendation samples in $D^{rec}$ to mitigate SSB in $D^{ad}$.
\section{Methodology}\label{sec:method}
Fig.~\ref{fig:framework} shows two stages of Rec4Ad in deployment: \textbf{data augmentation} and \textbf{disentangled representation learning}, which constructs and leverages recommendation (rec) samples, respectively.



\subsection{Data Augmentation}
\subsubsection{Retrieving Recommendation Samples} To ensure user experience, the percentage of ad impressions in all impressions is usually limited to a low threshold. Typically, we have $|D^{ad}| \ll |D^{rec}|$, making it intractable to consume entire $D^{rec}$ owing to multiplied training resources. Moreover, not all rec samples are useful for enhancing ads model due to difference between $\mathbf{\mathcal{A}}$ and $\mathbf{\mathcal{I}}$. Thus we retrieve rec samples that are closely related to advertising system. 

Let $I(a) \in \mathbf{\mathcal{I}}$ denote the item that $a \in \mathbf{\mathcal{A}}$ is advertised for. We define an item set $\mathbf{\mathcal{I}}^{ad}$ containing all items with related ads:
\begin{equation}
    \mathbf{\mathcal{I}}^{ad} = \{i\in \mathbf{\mathbf{I}}|\exists a \in \mathbf{\mathcal{A}},I(a)=i\}.
\end{equation}

We discard rec samples whose items fail to occur in $\mathbf{\mathcal{I}}^{ad}$, since they fail to provide complementary impressions that relate to any ad of interest. In this way, we retrieve a subset of rec samples:
\begin{equation}
    D^{\hat{rec}} = \{(u,i,c,y) \in D^{rec}|i\in \mathbf{\mathcal{I}}^{ad}\}.
\end{equation}

\subsubsection{Pseudo Sample Mapping} 
The reasons to map retrieved rec samples to pseudo ad samples are two-fold. First, it allows rec samples and ad samples to have uniform input format, which facilitates efficient feature joining and batch processing. Second, pseudo samples scattering in the  $\mathbf{\mathcal{U}} \times \mathbf{\mathcal{A}}$ space help learn a CTR prediction model for ads, compared with the $\mathbf{\mathcal{U}} \times \mathbf{\mathcal{I}}$ space.

In Fig.~\ref{fig:framework}, we maintain an item-ads index where key is item $i$ and values are their related ads $A(i) = \{a\in\mathbf{\mathcal{A}}|I(a) = i\}$. To select an ad from $A(i)$, we do not take their impressions into consideration, which avoids introducing selection bias in the advertising system. Instead, we adopt a recent-$K$-random strategy. We randomly select an ad $a'$ from most recent $K$ ads related to $i$, where $K$ is a fixed hyper-parameter. After mapping, we obtain the set of pseudo samples:
\begin{equation}
    D^{\tilde{rec}} = \{(u,a',c,y)|(u,i,c,y) \in D^{\hat{rec}}, a' \in A(i)\},
\end{equation}


\subsection{Disentangled Representation Learning}
\subsubsection{Original Representation}
we embed raw features of sample $(u,a,c,y)$ into low-dimensional vectors $[e(u),e(a),e(c)]$.
Operations like attention mechanism~\cite{zhou2018deep} are further employed to aggregate embeddings of user behavior sequences. We concatenate these results together to obtain intermediate representation $\mathbf{e}$. 

Batch Normalization (BN)~\cite{ioffe2015batch} is  commonly used in training of industrial CTR models~\cite{sheng2021one} to stabilize convergence. It calculate statistics over training data for normalization during serving. However, when incorporating $D^{\tilde{rec}}$ into training, BN statistics are calculated based on ad and rec samples but only used to normalize ad samples during online serving. The distribution discrepancy between two kinds of samples weakens the effectiveness of BN.

To deal with this problem, we design source-aware BN (SABN), which adaptively normalize samples according to their sources. Let $s$ indicate which kind the sample is, SABN works as follows:
\begin{equation}
    SABN(e,s) = \gamma_s\frac{e-\mu_s}{\sqrt{\sigma_s^2+\epsilon}} + \beta_s, s \in \{ad, rec\},
    \vspace{-2mm}
\end{equation}
where $\gamma_s, \beta_s, \mu_s, \sigma_s^2$ are source-specific parameters for normalization. Then we feed the normalized representation $\mathbf{e'}$ into MLP (Multi-Layer
Perception) layers for a compact representation $\mathbf{x}$ that captures feature interactions among user, ad, and context. We add superscripts on representations (e.g., $\mathbf{x}^{ad}$/$\mathbf{x}^{rec}$) to denote its source.


\subsubsection{Alignment} Since users are usually unaware of the difference between ad and rec impressions, their click decisions can be assumed independent of underlying systems, which are commonly determined by their interest. To identify user-item interest $\mathbf{\Gamma}$ behind click decision, we propose to extract invariant representations shared between $D^{ad}$ and $D^{\tilde{rec}}$. In other words, samples in $D^{ad}$ and $D^{\tilde{rec}}$ should be indistinguishably distributed in the invariant representation space. To achieve this goal, we first apply projection layers over original representations of ad and rec samples:
\begin{equation}
    \mathbf{x}^{s}_{inv} = MLP^{s}_{inv}(\mathbf{x}^{s}) \in \mathcal{R}^{d}, s \in \{ad, rec\}
\end{equation}

A direct method to align $\{\mathbf{x}^{ad}_{inv}\}$ and $\{\mathbf{x}^{rec}_{inv}\}$ is minimizing their Wasserstein or MMD distribution distance~\cite{cuturi2014fast,gretton2012kernel}. However, these metrics are computationally inefficient and hard to estimate accurately over mini-batches. Instead, we train a sample discriminator $H$ to implicitly align them in an adversary way. Particularly, $H$ is a binary classifier that predicts whether the sample is from $D^{ad}$ or $D^{\tilde{rec}}$ based on $\mathbf{x}_{inv}$.
Optimized with cross entropy loss, $H$ aims to distinguish two kinds of samples as accurate as possible:
\begin{equation}
\begin{aligned}
    &\hat{s} = Sigmoid(MLP_H(\mathbf{x}_{inv}^s)), s\in \{ad, rec\},\\
    &L_{A} = -\sum_{D^{ad}}log(\hat{s})-\sum_{D^{\tilde{rec}}}log(1-\hat{s}).
\end{aligned}
\end{equation}
While $H$ tries to minimize $L_{A}$ during training, neural layers generating invariant representations aim to make $\{\mathbf{x}^{ad}_{inv}\}$ and $\{\mathbf{x}^{rec}_{inv}\}$ indistinguishable as much as possible, i.e., maximize $L_{A}$. To train these two parts simultaneously, we insert a gradient reverse layer (GRL)~\cite{ganin2015unsupervised} between $\mathbf{x}_{inv}$ and the discriminator. In forward propagation, GRL acts as an identity transformation. In backward propagation, it reverses gradients from subsequent layers:
\begin{equation}\label{eq:grl}
    \begin{aligned}
        &\text{Forward}:  GRL(\mathbf{x}_{inv}) = \mathbf{x}_{inv},\\
        &\text{Backward}: \frac{\partial L_A}{\partial \mathbf{x}_{inv}} = -\alpha \frac{\partial L_A}{\partial GRL(\mathbf{x}_{inv})},
    \end{aligned}
\end{equation}
where $\alpha$ controls the scale of reversion. In this way, we tightly align $\{\mathbf{x}^{ad}_{inv}\}$ and $\{\mathbf{x}^{rec}_{inv}\}$ in the invariant representation space.

\subsubsection{Decorrelation} To separate system-specific confounders from original representation, we apply another set of projection layers: \begin{equation}
    \begin{aligned}
    \mathbf{x}^{s}_{con} &= MLP^{ad}_{con}(\mathbf{x}^{s})  \in \mathcal{R}^{d}, s \in \{ad, rec\}.
    \end{aligned}
\end{equation}
If without explicit constraints, $\mathbf{x}_{con}$ could still contain information shared across systems and prevent us from handling confounders specific to the ad system. To this end, we propose to add regularizations to further disentangle $\mathbf{x}_{con}$ and $\mathbf{x}_{inv}$.

Borrowing the idea that disentangled representations avoid encoding variations of each other~\cite{cheung2014discovering,cogswell2015reducing}, we penalize the cross-correlation between two sets of representations. Specifically, let $\mathbf{p}_i$ denote the in-batch vector of $i$-th dimension of $\mathbf{x}_{inv}$ and $\mathbf{q}_j$ denote that of $j$-th dimension of $\mathbf{x}_{con}$, their Pearson correlation can be calculated as:
\begin{equation}
\begin{aligned}
    &Cov(\mathbf{p}_i, \mathbf{q}_j)) = [\mathbf{p}_i-\mathbf{\bar{p}_i}]^\top[\mathbf{q}_j-\mathbf{\bar{q}_j}],\\
    &\Upsilon(\mathbf{p}_i, \mathbf{q}_j) = \frac{Cov(\mathbf{p}_i, \mathbf{q}_j)}{\sqrt{Cov(\mathbf{p}_i, \mathbf{p}_i)}},
\end{aligned}
\end{equation}
where $\mathbf{\bar{p}_i}$ and $\mathbf{\bar{q}_j}$ denote in-batch mean of each dimension. Thus the objective of the decorrelation module are based on correlations of every pair of dimension cross $\mathbf{x}_{inv}$ and $\mathbf{x}_{con}$:
\begin{equation}
    \begin{aligned}
        L_{D} = \sum_{i=1}^{d}\sum_{j=1}^{d} [\Upsilon(\mathbf{p}^{ad}_i, \mathbf{q}^{ad}_j)^2 + \Upsilon(\mathbf{p}^{rec}_i, \mathbf{q}^{rec}_j)^2].
    \end{aligned}
\end{equation}

By optimizing $L_{D}$, $\mathbf{x}_{con}$ are encouraged to capture residual information independent from $\mathbf{x}_{inv}$, i.e., system-specific confounders $\mathbf{\Delta_{ad}}$/$\mathbf{\Delta_{rec}}$ that are discarded by the alignment module.

\subsection{Prediction} We reconstruct final representation based on disentangled representations to predict CTR. Previous studies show that non-causal associations also potentially contribute to prediction accuracy~\cite{zhang2021causal,si2022model}, motivating us to consider $\mathbf{x}_{con}$ in reconstruction instead of directly ignoring it. For simplicity, we use the concatenation operator:
\begin{equation}
    \mathbf{x}^{s}_{new} = \mathbf{x}^{s}_{inv} \oplus \mathbf{x}^{s}_{con}, s \in \{ad, rec\}.
\end{equation}
With $\mathbf{x}_{new}^s$, we make predictions for ad samples and pseudo samples with source-aware layers, where cross entropy loss $L_C$ is optimized:
\begin{equation}
\begin{aligned}
    &\hat{y}^{s} = Sigmoid(MLP^{s}_{pred}(\mathbf{x}^{s}_{new})),  s \in \{ad, rec\}\\
    &L_C = \sum_{D^{ad} \cup D^{\tilde{rec}}}[-ylog(\hat{y}^{s})-(1-y)log(1-\hat{y}^{s})].
\end{aligned}
\end{equation}
Thus the objective function of Rec4Ad consists of the CTR prediction loss, the alignment loss and the decorrelation loss:
\begin{equation}\label{eq:loss}
    L = L_{C} + \lambda_{1}L_{A} + \lambda_{2}L_{D}.
\end{equation}

\section{Experiments}
\subsection{Experimental Setup}\label{sec:setup}
\noindent\textbf{Taobao Production Dataset.}
We construct the dataset based on impression logs in two weeks of 2022/06 from Taobao advertising system and recommendation system. We use data of the first week for training, which contains ad and rec impressions collected under regular policy. The data of the next week are ad impressions collected under random policy of a small traffic following~\cite{yuan2019improving,uniform20}, which is used to evaluate model performance against SSB. The training dataset contains 1.9 billion ad samples and 0.6 billion rec samples after retrieval, covering 0.2 billion users.  The test dataset contains 18.9 million ad samples and 10.3 million users. 

\noindent\textbf{Baselines.} Rec4Ad is compared with following baselines.
\begin{compactitem}
    \item \textbf{Base}. We adopt DIN~\cite{zhou2018deep} as the vanilla model which does not account for SSB.
    \item \textbf{DAG}. The Data-Augmentation (DAG) method directly merges rec samples and ad samples to train the base model.
    \item \textbf{IPS}~\cite{joachims2017unbiased,pmlr-v48-schnabel16}. It eliminates SSB  by re-weighting samples with inverse propensity of ad impression.
    \item \textbf{IPS-C}~\cite{bottou2013counterfactual} It adds max-capping to IPS weight so that its variance can be reduced. 
    \item \textbf{IV}~\cite{si2022model}. It employs user behaviors outside current system as instrumental variables for model debiasing.
\end{compactitem}

\noindent\textbf{Metrics.}
For ranking ability, we use the standard \textbf{AUC} (Area Under
the ROC Curve) metric for evaluation~\cite{lian2018xdeepfm,guo2017deepfm}. A higher AUC indicates better ranking performance. \textbf{In practice, absolute improvement of AUC by 0.001 on the production dataset is considered significant, which empirically leads to an online lift of 1\% CTR}. For calibration, we evaluate models with the \textbf{ECE}~\cite{yan2022scale} metric. We first equally partition the $pctr$ range [0,1] into $K$ buckets $B_1,...,B_K$. ECE can be calculated as follows:
\begin{equation}
    \text{ECE} = \frac{1}{|D|}\sum_{k=1}^{K}|\sum_{i=1}^{|D|}(y_i-\hat{y}_i)\:\vmathbb{1}(\hat{y}_i\in{B_k})|,
\end{equation}
where $\vmathbb{1}(\hat{y}_i\in{B_k})$ equals 1 only if $\hat{y}_i\in{B_k}$ else 0. $K$ is set to 100. A lower ECE here indicates better calibration performance.

\noindent\textbf{Implementation}
The feature embedding size is 16. We use Adam optimizer~\cite{kingma2014adam} with initial learning rate $0.001$. The batch size is fixed to 6000. In data augmentation, we consider the most recent 3 ads for pseudo sample mapping. The dimensions $d$ of $\mathbf{x}_{inv}$ and $\mathbf{x}_{con}$ is 128. The ratio $\alpha$ of gradient reverse layer in Eq.~(\ref{eq:grl}) is 0.1. $\lambda_1$ and $\lambda_2$ for the alignment and the decorrelation loss in Eq.~(\ref{eq:loss}) is 0.005 and 0.5. For tests of significance, each experiment is repeated 5 times by random initialization and we report the average as results.

\subsection{Experimental Results}
\begin{table}[!htbp]
	\centering
	\caption{Performance Comparison of Rec4Ad and baselines. The symbol * indicates the improvements over baselines are
significant with p-value < 0.01 by t-test. We omit standard deviations since they are all no more than $2\times10^{-4}$.}\label{tab:performance}	
    \vspace{-1.5mm}
	\resizebox{.75\columnwidth}{!}{
		\begin{tabular}{c|cc|cc}
			\toprule
			Method & AUC & Impv. & ECE & Impv.\\
			\midrule
			Base& 0.6778 & - & 0.0007 & -\\
			DAG& 0.6724 & -0.0054 & 0.0032 & -0.0025\\
			IPS& 0.6618& -0.0160& 0.0023 & -0.0016\\
            IPS-C& 0.6783& +0.0005& 0.0015 & -0.0008\\
            IV& 0.6790& +0.0012& 0.0009  & -0.0002 \\
            \midrule
            Rec4Ad& \textbf{0.6805*}& \textbf{+0.0027}& \textbf{0.0002} & \textbf{+0.0005}\\
			\bottomrule
		\end{tabular}
	}
 \vspace{-3mm}
\end{table}

\subsubsection{Overall Performance} From Table~\ref{tab:performance}, we find that Rec4Ad significantly performs better than all baselines. Specifically, it outperforms Base in terms of AUC by 0.0027 and outperforms the state-of-the-art IV by 0.0015. This demonstrates the effectiveness of our proposed framework in handling SSB. By dissecting confounders and user-item interest for enhanced representations, it works well over the inference space. Moreover, Rec4Ad successfully maintains even slightly better model calibration than Base, which also verifies its suitability for ads CTR prediction. We also observe that DAG performs worse than baseline both in AUC and ECE. The reason is ad samples and rec samples present different feature distributions and label distributions. Naive data augmentation actually amplifies the distribution discrepancy between training and inference.  The original IPS yields worst AUC, while IPS-C with max-capping achieves higher AUC than Base. We attribute this phenomenon to high variance in estimation of propensity score. We also notice that ECE of IPS and IPS-C are all larger than $10^{-3}$, which verifies that sample re-weighting could change label distribution and result in calibration issues of ads CTR prediction. 

\subsubsection{Performance on Different Ad Groups}
\begin{table}[!htbp]
	\centering
	\caption{Comparison among Rec4Ad and three competitive baselines on  ad groups with different impression ratios.}\label{tab:group-performance}		
	\resizebox{.85\columnwidth}{!}{
		\begin{tabular}{c|c|cc|cc}
			\toprule
			 Group & Method & AUC & Impv. & ECE & Impv.\\
			\midrule
            \multirow{4}{*}{$G_{top}$} & Base & 0.6741& -& \textbf{0.0002} & -\\
            & IPS-C& 0.6736& -0.0005& 0.0004  & -0.0002 \\
            & IV& 0.6743& +0.0002& \textbf{0.0002}  & 0 \\

            & Rec4Ad& \textbf{0.6757*}& \textbf{+0.0016}& \textbf{0.0002} & 0\\   
            \midrule
            \multirow{4}{*}{$G_{bottom}$} & Base & 0.6625& -& 0.0021 & -\\
            & IPS-C& 0.6648& +0.0023& 0.0018  & +0.0003 \\
            & IV& 0.6734& +0.0009& 0.0021  & 0 \\

            & Rec4Ad& \textbf{0.6654*}& \textbf{+0.0029}& \textbf{0.0010*} & \textbf{+0.0011}\\ 
            \bottomrule
		\end{tabular}
	}
\end{table}

In Section~\ref{sec:influence}, we show that SSB leads model to perform badly on ads with low impression ratios. To validate whether Rec4Ad mitigates such influence, we compare Rec4Ad and three competitive baselines on specific ad groups. We sort ads in descending IR as defined in Eq.~(\ref{eq:ir}), where the top 25\% are selected as $G_{top}$ representing ads with enough impressions and the bottom 25\% are selected as $G_{bottom}$  containing ads that are less represented in the training data.

Table~\ref{tab:group-performance} shows that Rec4Ad achieves best ranking and calibration performance on both $G_{top}$ and $G_{bottom}$. The improvements over Base are greater on $G_{bottom}$ with AUC increased by nearly 0.003 and ECE reduced by 0.001. Thus we conclude that Rec4Ad succeeds in mitigating SSB and boosts model performance on those long-tail ads. We also observe an interesting seesaw phenomenon about IPS-C, which also greatly improves metrics on $G_{bottom}$ but yields worse performance on $G_{top}$ compared with Base. It is because IPS-C explicitly imposes higher weights for samples with low-IR ads and lower weights for those with high-IR ads. By contrast, Rec4Ad exhibits its superiority that improvements on $G_{bottom}$ are achieved without the cost of degraded performance on $G_{top}$.

\subsubsection{Ablation Study}
\begin{figure}[!htbp]
	\centering	
	\includegraphics[width=0.47\columnwidth]{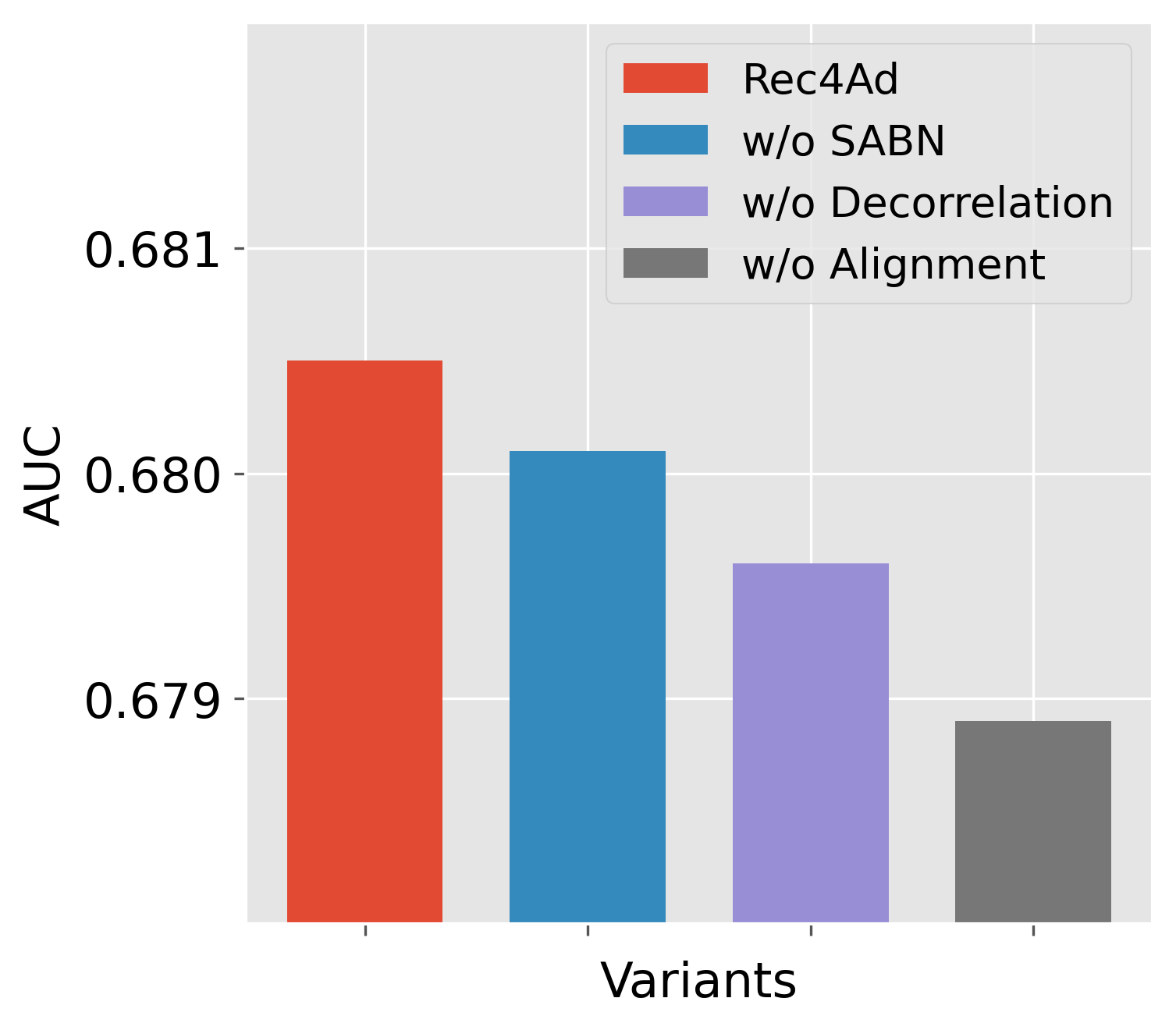}    
	\includegraphics[width=0.44\columnwidth]{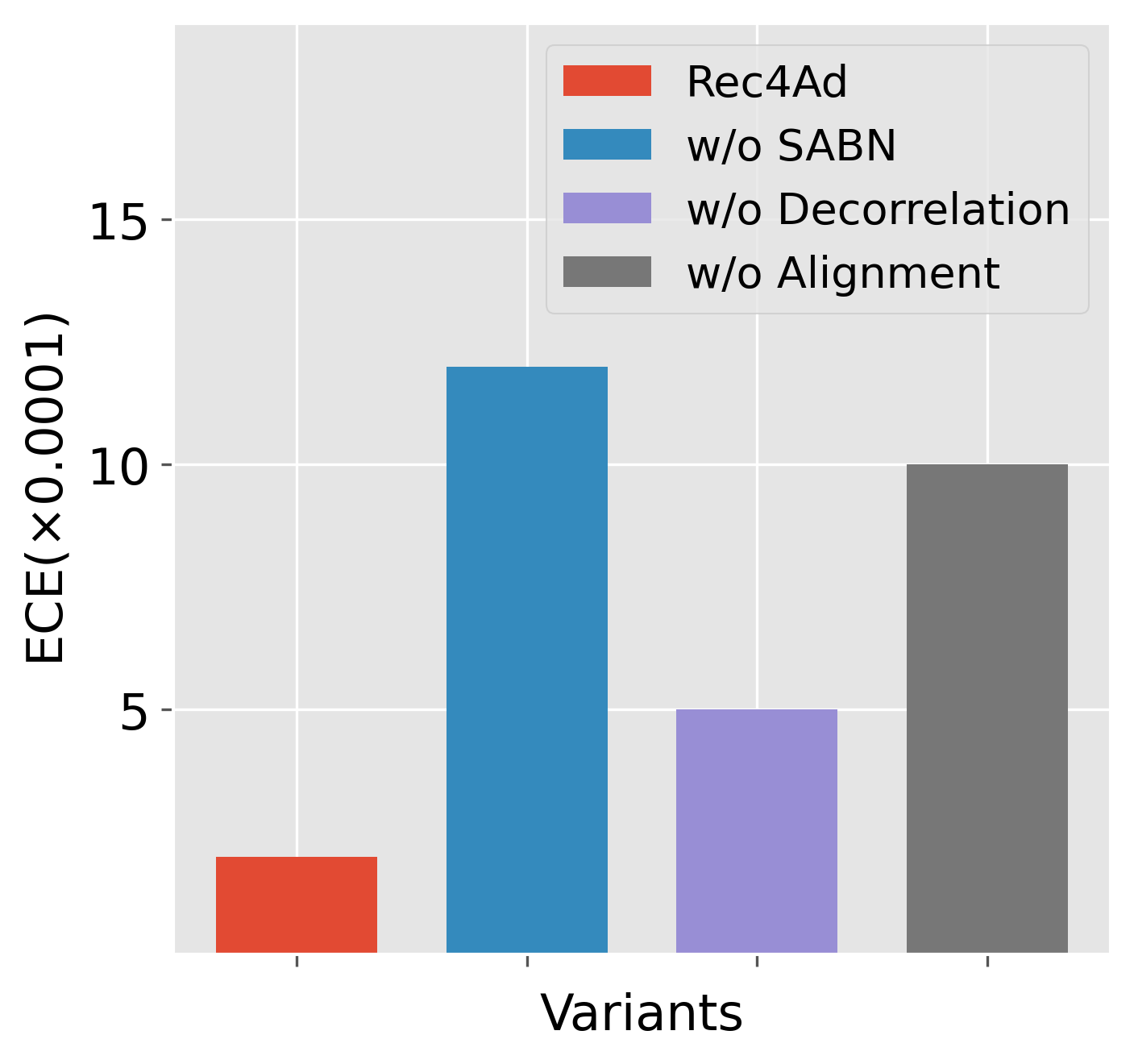}   
	\caption{Comparison among Rec4Ad and its variants.}\label{fig:ablation}
\end{figure}

We analyze the effect of key components in Rec4Ad by comparing it with variants which remove SABN, the alignment module, and the decorrelation module, respectively.

Fig.~\ref{fig:ablation} shows that after removing SABN, model calibration experiences an obvious degeneration. The reason is that representations of Rec and Ad samples are with different distributions, making it difficult to normalize them with shared BN parameters and leading to mis-scaled network activations as well as badly-calibrated predictions. Furthermore, we find that AUC even drops under $0.679$ after removing the alignment module, validating that the alignment regularization is critical for co-training with ad and rec samples. It allows Rec4Ad to extract shared user-item interest behind user clicks and eliminate system-specific confounders from this part. The decorrelation module is also shown effective since the variant without this component performs worse than the default version. It is because splitting non-causal correlations alone in enhanced representations also potentially contributes to accurate predictions~\cite{zhang2021causal,si2022model}.

\subsubsection{Study on Disentangled Representations} 
\begin{figure}[!htbp]
	\centering	
	\includegraphics[width=0.49\columnwidth]{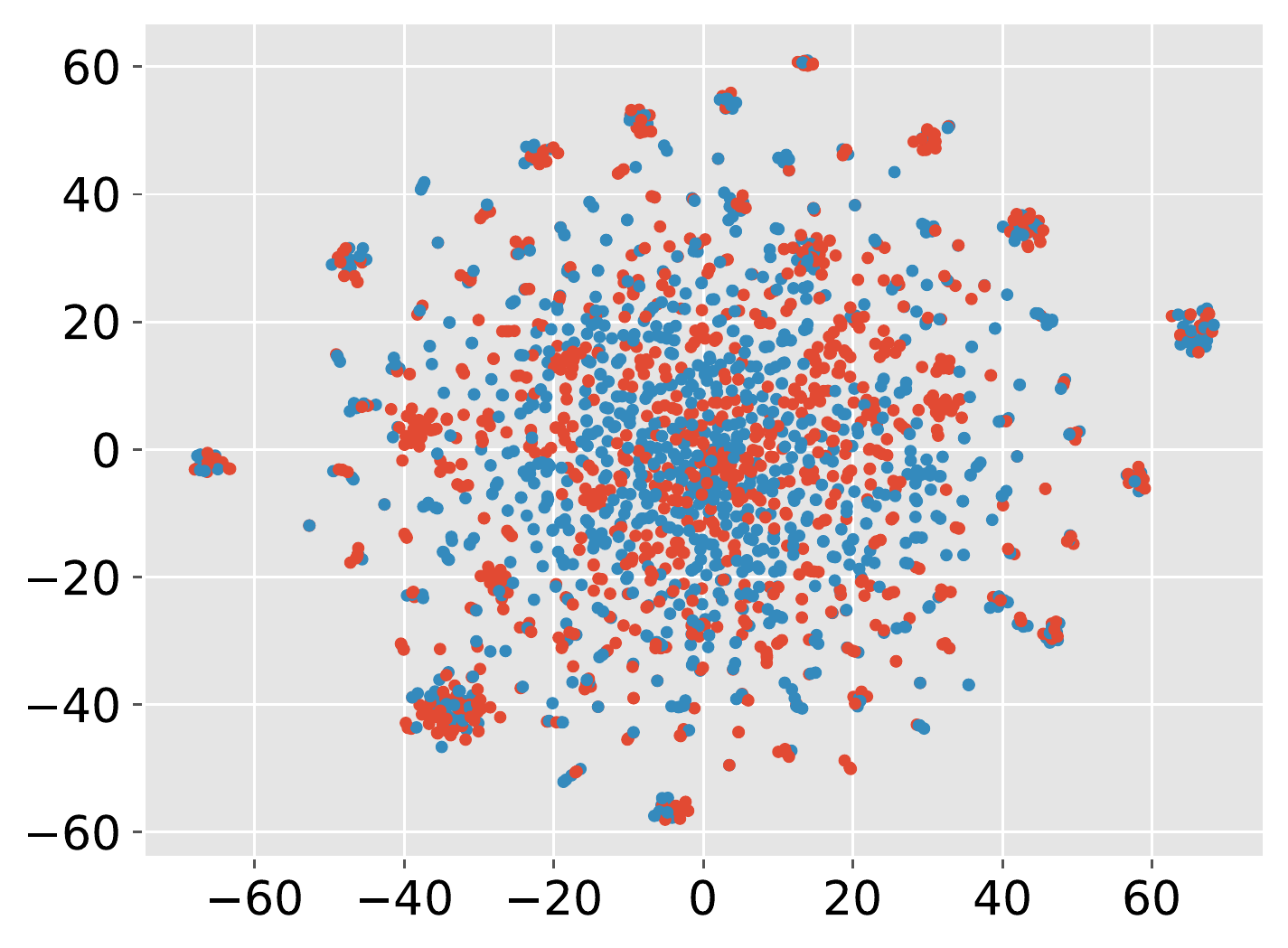}    
	\includegraphics[width=0.49\columnwidth]{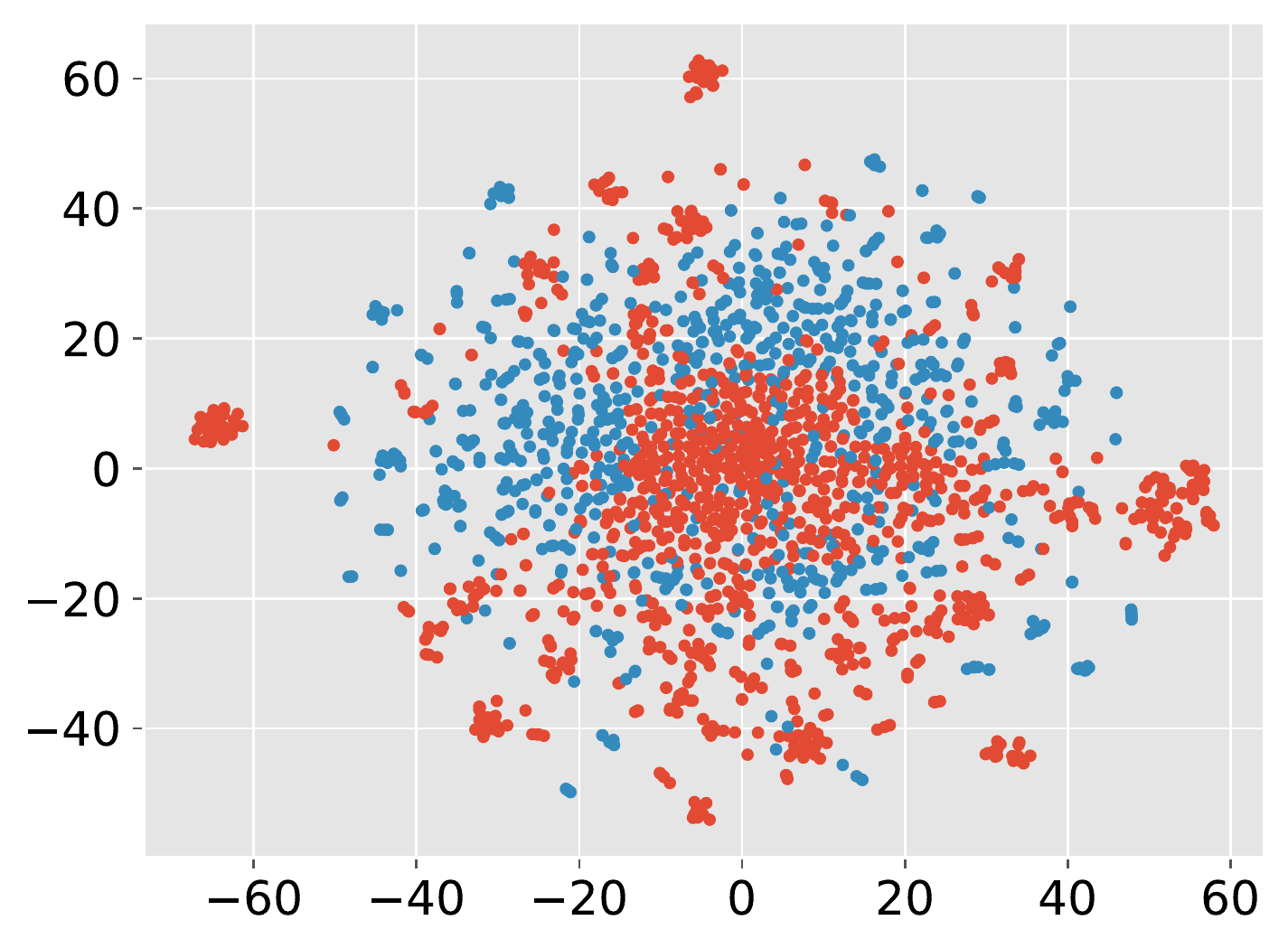}   
	\caption{T-SNE visualization of $\mathbf{x}_{inv}$ (Left) and $\mathbf{x}_{con}$ (Right) for ad samples (red) and rec samples (blue).}\label{fig:align_decor}
\end{figure}

\begin{figure}[!htbp]
	\centering	
	\includegraphics[width=0.7\columnwidth]{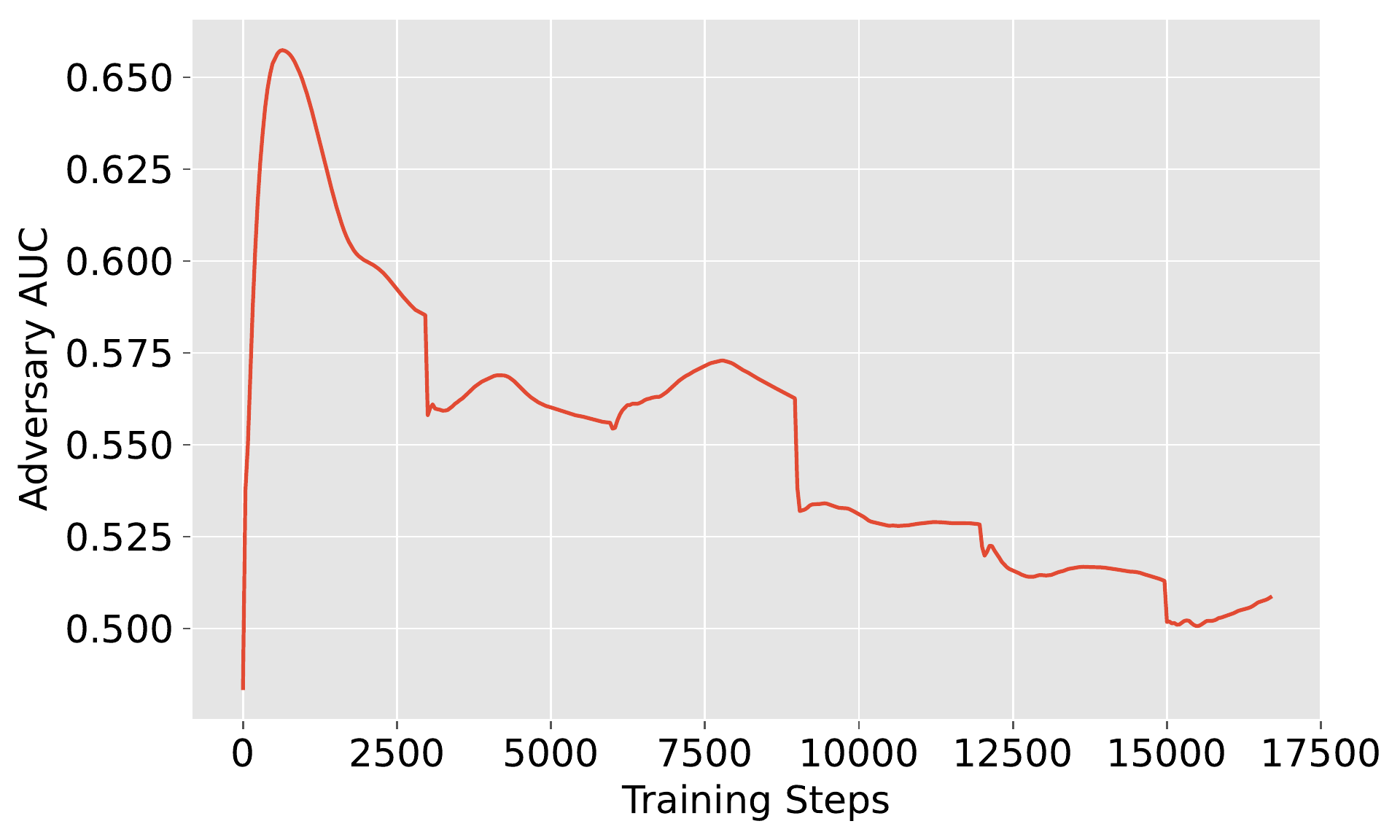}    
	\caption{Training AUC of the adversary sample classifier $H$.}\label{fig:dann_auc}
\end{figure}


As $\mathbf{x}_{inv}$ and $\mathbf{x}_{con}$ are expected to capture system-invariant and system-specific factors respectively, we aim to investigate their distributions over ad and rec samples. We randomly sample a hybrid batch and visualize learned representations using t-SNE~\cite{van2008visualizing}. As shown in Fig.~\ref{fig:align_decor}, there is no significant difference between $\mathbf{x}_{inv}$ for ad and rec samples, suggesting the captured invariance.  When it comes to $\mathbf{x}_{con}$, we observe that ad samples and rec samples are mostly separated in different areas, indicating that this representation extracts system-specific factors from the training data, which we believe stems from the difference in their selection mechanisms. 

In the alignment module, we employs an adversary sample discriminator $H$ to distinguish $\mathbf{x}_{inv}^{ad}$ and $\mathbf{x}_{inv}^{rec}$. Thus the classification performance can be used as an effective proxy to quantitatively evaluate the goodness of $\mathbf{x}_{inv}$. Fig.~\ref{fig:dann_auc} illustrates the adversary AUC during training. We observe that it increases at the early stage due to optimization of $H$. Then AUC gradually decreases as the training goes on, indicating Rec4Ad tries to generate representations that confuse $H$. Near the end of training, AUC converges to 0.5, which means ad and rec samples are indistinguishable on $\mathbf{x}_{inv}$.

\subsection{Online Study}
\begin{table}[!htbp]
	\centering
	\caption{Key business metrics of online A/B Test.}\label{tab:online-ab}		
	\resizebox{.75\columnwidth}{!}{
		\begin{tabular}{c|cc|cc}
			\toprule
            \multirow{2}{*}{Scene} & \multicolumn{2}{c|}{Overall} & \multicolumn{2}{c}{Long-Tail}\\
            \cline{2-5}
            & CTR & RPM & CTR & RPM\\
            \midrule
			Homepage& +6.6\%& +2.9\%& +12.6\% & +9.3\%\\
            Post-Purchase & +3.0\% & +2.6\% & +3.6\% & +0.8\%\\
			\bottomrule
		\end{tabular}
	}
\end{table}

We conduct online A/B Test between Rec4Ad and production baseline from July 1 to July 7 of 2022, each with 5\% randomly-assigned traffic. Two key business metrics are used in evaluation: Click-Through Rate (CTR) and Revenue Per Mille (RPM), which corresponds to user experience and platform revenue, respectively. As shown in Table~\ref{tab:online-ab}, Rec4Ad achieves substantial gains in two largest scenes of Taobao display advertising business, \emph{Homepage} and \emph{Post-Purchase}, demonstrating considerable business value of Rec4Ad. For long-tail ads with few impressions. Rec4Ad achieves up to 12.6\% and 3.6\% lift of CTR in two scenes, which are larger than the overall lift. Above results verify that Rec4Ad effectively mitigates SSB and brings solid online improvements. It has been successfully deployed in production environment to serve the main traffic of Taobao display advertising system since July of 2022.
\section{Conclusion}
In this paper, we propose a novel framework which leverages
\textbf{Rec}ommendation samples to help mitigate sample selection bias \textbf{F}or \textbf{Ad}s
CTR prediction (\textbf{Rec4Ad}).  Recommendation samples are first retrieved and mapped to pseudo samples. Ad samples and pseudo samples are jointly considered in learning disentangled representations that dissect system-specific confounders brought by selection mechanisms and system-invariant user-item interest. Alignment and decorrelation modules are included in above architecture. When deployed in Taobao display advertising system, Rec4Ad achieves substantial gains in key business metrics, with a lift of up to +6.6\% CTR and +2.9\% RPM.


\balance 
\bibliographystyle{ACM-Reference-Format}
\bibliography{reference}

\end{document}